\documentclass[12pt]{iopart}

\usepackage{color}

\usepackage{ifsym}  
\usepackage{amssymb}  
\usepackage{textcomp}
\usepackage[breaklinks,hidelinks]{hyperref} 
\hypersetup{colorlinks=true,linkcolor=blue,citecolor=blue,urlcolor=blue}

\usepackage{graphicx}
\usepackage{bm}
\usepackage{url}  
\usepackage{cite} 
\usepackage{iopams}
\begin{document}


\title[The formation of intermediate layers in covered Ge/Si heterostructures\dots]{The formation of intermediate layers in covered Ge/Si heterostructures with low-temperature quantum dots: a study using high-resolution transmission electron microscopy and Raman spectroscopy}

\author{ Mikhail~S~Storozhevykh,$^1$ 
	Larisa~V~Arapkina,$^1$\footnote[1]{E-mail:~arapkina@kapella.gpi.ru}
	Sergey~M~Novikov,$^2$
	Valentyn~S~Volkov,$^2$
	Oleg~V~Uvarov$^1$
	and
	Vladimir~A~Yuryev$^1$\footnote[6]{E-mail:~vyuryev@kapella.gpi.ru}
}


\

\address{$^1$ A.\,M.\,Prokhorov General Physics Institute of the Russian Academy of Sciences, 38 Vavilov Street, Moscow, 119991, Russia}
\ead{storozhevykh@kapella.gpi.ru}

\

\address{$^2$ Center for Photonics and 2D Materials, Moscow Institute of Physics and Technology, 9 Institutskiy per., Dolgoprudny, Moscow Region, 141707, Russia}

\begin{abstract}
The method of software analysis of high-resolution TEM images using the peak pairs algorithm in combination with Raman spectroscopy was employed to study lattice deformations in Ge/Si(001) structures with low-temperature Ge quantum dots. 
It was found that the stresses do not spread in a thick Si layer above quantum dots, but completely relax via the formation of a thin boundary layer of mixed composition. 
However, intermixing of Ge and Si is absent beneath the Ge layer in samples with a Ge coverage of 10~{\AA}. Besides intermixing was not observed at all, both beneath and above the Ge layer, in samples with a Ge coverage of 6~{\AA} or less. 
This may be due to the predominance of Ge diffusion into the Si matrix from the \{105\} facets of Ge huts, not from the Ge wetting layer, at low temperatures of the Ge/Si structure deposition. 
The critical thickness of Si coverage at which the intense stress-induced diffusion takes place is determined to lie in the range from 5 to 8 nm.
\end{abstract}

\vspace{2pc}


\maketitle

\section{\label{sec:intro}Introduction}

Quantum dots of different type and compound are widely used in optoelectronic-related fields as a core ingredient of light detectors 
\cite{Dual-band_infrared_imaging_ccolloidal_QD,A_monolithically_integrated_plasmonic,Quantum_Dots-in-a-Well_FPA},
emitters 
\cite{Emission_from_quantum-dot_microcavities,Asymmetrically_strained_quantum_dots,What_future_for} 
and even lasers 
\cite{Lasing_from_glassy_Ge,Quantum_cascade_lasers}. 
And stresses arising in them play very important role in many cases 
\cite{Correlating_photoluminescence,Asymmetrically_strained_quantum_dots,Electron_spatial_localization_tuned_by_strain,A_new_route_toward_light_emission}.

A detailed study of the morphology of semiconductor nano-heterostructures is an extremely important task, the solution of which is always limited by the level of development of methods for studying the crystal lattice of a material with the highest possible spatial resolution.
One of the most common methods enabling one to obtain a direct image of the crystal internal structure down to individual atoms is high-resolution transmission electron microscopy (HR~TEM).
Studies of lattice strains and defects in stressed semiconductor structures with thin layers and nanoclusters using this technique is of great interest
\cite{Quantitative_strain_analysis,A_new_route_toward_light_emission,Strain_relaxation_in_self-assembled}.

A lot of works concerning strains in Ge/Si structures with quantum dots (QDs) published in the last decades are devoted to QDs grown at relatively high substrate temperatures \cite{Strain&composition_Ge/Si,Lattice_deformation@Ge-QD,37_Si/Ge_Raman_strain&intermixing,39_QD_strain&composition,41_Gigantic_diffusion_SiGe,44_Polarized_Raman_Ge-Si-QD,45_Raman_Si-Ge_QD_Abstreiter,46_Strain&composition_Si-Ge_QD}. 
And most of experimental works include Raman spectroscopy \cite{our_Raman_en,37_Si/Ge_Raman_strain&intermixing,38_SiGe_lateral_ordering,39_QD_strain&composition,40_Raman&Capacitance_Si/Ge_QD,41_Gigantic_diffusion_SiGe,42_Morphology&optical_properties_Ge_nano-films,43_Phonons_GeSi-QDs,44_Polarized_Raman_Ge-Si-QD,45_Raman_Si-Ge_QD_Abstreiter,46_Strain&composition_Si-Ge_QD} 
since it is a rapid and nondestructive method imposing no special requirements on processing of a sample. Unfortunately, this method has a serious restriction: its spatial resolution is too low to discriminate between nano-sized objects even using confocal microscopes that gives no way to directly separate the contribution from thin layers of a multilayer structure. 
It becomes especially critical if strains and composition of the heterostructure vary gradually. 
The combination of the two above-mentioned methods is an effective approach allowing one to successfully study complex multilayer heterostructures containing layers of variable composition. 

In the present work, we attempt to shed light on intrinsic formation mechanisms of a Si/Ge heterointerface during covering of low-temperature Ge QDs by Si using molecular-beam epitaxy (MBE). For this purpose Ge/Si(001) heterostructures with low-temperature ($T_{\rm Ge}\approx360${\textcelsius}) self-organizing Ge QDs, which are of special interest due to the possibility of easy integration of their formation process with the standard CMOS process of the VLSI production, were chosen as an object of the research. High density of low-temperature Ge/Si(001) QDs ($\sim 6\times 10^{11}$~cm$^{-2}$ \cite{classification}) is an advantage that enables considering these structures for employing in optoelectronics, whereas the band structure of hole states promises their efficient utilization for detecting of electromagnetic radiation in mid-wave infrared (3 to 5 \textmu m)  \cite{Gerasimenko_Si-mat_nanoelectr,Dvurech_Brudnyi}. 

As it was shown earlier \cite{Yur_JNO,Growing_Ge_hut-structure}, 
when a thin layer of Ge is deposited by MBE on Si, stress occurring due to the greater lattice parameter of Ge relaxes already in the wetting layer with the growth of $(M\times N)$-patches separated by deep trenches of dimer-row vacancies penetrating the wetting layer to the full depth up to the silicon substrate and shallower trenches of dimer vacancy lines reaching a depth of 1 or sometimes 2 monolayers. In the first two monolayers of the Ge wetting layer, the lattice parameter of patches $a_{[110]}$ measured using a scanning tunnelling microscope (STM) was $\sim 3.8$~\AA, which corresponds to that of the unstrained Si lattice, and its value reached 4~\r{A} in the fourth and fifth monolayers (the measured values of $a_{[110]}$ sometimes exceeded $\sim 4.2$~\AA), i.e. the upper layers of Ge/Si(001) wetting layer patches have the lattice parameter of unstrained Ge and are completely relaxed \cite{Yur_JNO,Growing_Ge_hut-structure}.
Nikiforov \textit{et~al.} obtained similar results using the recording diffractometry during the growth of Ge/Si(001) structures by MBE \cite{Nikiforov_Ge-Si_RHEED-oscillations,Nikiforov_Ge-Si_RHEED}; 
Teys concluded the same discussing possible nucleation mechanisms for Ge/Si(001) huts \cite{Teys_WL}. 
Our experiments have shown \cite{VCIAN-2012,CMOS-compatible-EMRS,Yur_JNO,Growing_Ge_hut-structure,initial_phase} 
that Ge huts nucleate on the topmost layers of wetting layer patches that are of sufficient sizes for formation of the 16-dimer nuclei structures \cite{CMOS-compatible-EMRS,Hut_nucleation,VCIAN2011} 
and have reached the thickness of 4 monolayers \cite{initial_phase}, 
i.e. hut nuclei arise on completely relaxed layers of Ge. 
Growing huts are faceted by \{105\} planes and also completely relaxed, i.e. have the lattice constant of undisturbed Ge. 

Yet, if we consider Ge clusters covered by Si layers that alters matters. 
In this case, either Ge is subject to compressive strain or Si above it becomes tensile strained.
The study of strains arising in multilayer Ge/Si heterostructures with QDs is important for any problems associated with the use of such structures in optoelectronics because the environment of QDs plays an important role in the formation of the band structure. The impact of a cap layer on the photoluminescence of InGaAs QDs was shown in 
\cite{Correlating_photoluminescence}.
Stresses in the vicinity of Ge layer cause tensile strains of the Si crystal lattice, which leads to the bending of the band structure with the formation of a potential well for electrons in the conduction band of Si near the QD \cite{Meyer}. 
Due to this, exciton forming and radiative recombination via indirect transitions in real space between hole states in a QD and electronic states in a bent conduction band of Si become possible. 
The energy of these transitions falls into the transparency range of the optical fibre (1.3 to 1.55 {$\mu$m) that arouse the great interest in terms of the use of Ge/Si structures as emitters and detectors of signal in fibre-optic communication lines. Recent advances in this field are associated with the use of heterostructures with Ge QDs enriched with interstitial defects \cite{Laser_Ge_QD,Laser-2_Ge_QD}. 
These defects form additional levels in the QD band gap, thereby enabling direct in real space transitions within the QD. 
However, even in this case, the potential wells for electrons in the conduction band of Si play a significant role, increasing the probability of electron tunnelling to the levels of defects in the QD. 
On the contrary, structures with the smallest potential well depth for electrons are preferred for use as mid-wave IR detectors. 
The absorption of IR radiation occurs through transitions of holes from the bound states of a QD to the valence band of silicon, and the recombination of electron-hole pairs is a negative effect contributing to emptying of hole states in QDs.


\section{\label{sec:exp}Samples, methods and equipment}

\begin{figure}[ht]
	\centering
	\includegraphics[width=.27\textwidth]{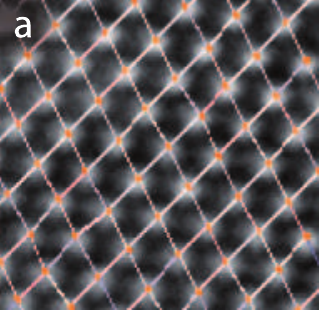}
	\includegraphics[width=.27\textwidth]{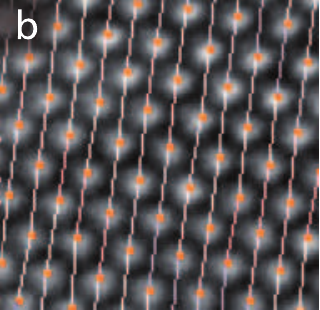}
	\includegraphics[width=.27\textwidth]{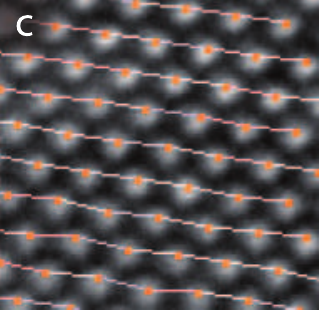}\\~\\
	\includegraphics[width=.4\textwidth]{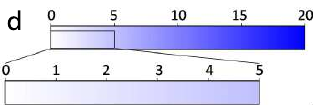}~
	\includegraphics[width=.4\textwidth]{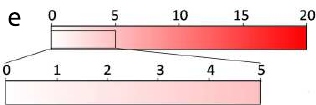}
	\caption{\label{fig:Colors}
		The fragment of HR TEM image of Si lattice processed with ``nearest neighbours'' template (a), ``growth direction'' template (b) and ``lateral direction'' template (c); the palette of correspondence between colour and strain values of inter-atomic bonds in percent: blue colour (d) corresponds to tensile strain, red one (e) represents compression strain.
	}
\end{figure}

The studied samples with layers of Ge QDs in the Si matrix were grown by molecular-beam epitaxy (MBE) in ultra-high vacuum (UHV) using an EVA\,32 chamber of an SSC\,2 surface-science centre (Riber) equipped with Ge and Si sources with the electron-beam evaporation. 
Czochralski-grown (100) oriented $p$-type silicon wafers of 50~mm in diameter 
($\varrho=$ 12~{\textohm}cm) 
were used in experiments.
The wafers were subjected to preliminary chemical treatments 
followed by annealing at the temperature of about {600\textcelsius} 
for removing adsorbates from the silicon surface 
lasted for 6 hours in the high-vacuum preliminary cleaning chamber. After that the wafers were moved into the UHV  MBE chamber where final deoxidation of the cleaned surface was made 
before the heterostructure growth, as a result of which the silicon oxide was completely removed and an atomically clean  Si(001) surface ready to MBE was obtained.
The removal of a silicon oxide layer was carried out by sample annealing at {800\textcelsius}
in a flux of Si atoms of $\lesssim 0.1$~\r{A}/s until a total amount of the deposited Si
reached 30~\AA; 2-minute stoppages
of Si deposition were made first twice after every 5~\r{A}
and then twice after every 10~\r{A} \cite{VCIAN-2012}.%
\footnote{More detailed information on the sample cleaning processes and the structure of the resultant Si(001) surface can be found, e.g., in Refs.~\cite{classification,our_Si(001)_en,phase_transition,CMOS-compatible-EMRS,VCIAN-2012,stm-rheed-EMRS}.}
Afterwards, a Si buffer layer of $\sim 100$ nm in thickness was grown on the prepared surface at the substrate temperature of {650\textcelsius}.
Layers of Ge QDs were grown at the temperature of {360\textcelsius}; the Ge coverage ($h_{\rm Ge}$) was varied in different samples from 6 to 10~{\AA} \cite{classification, CMOS-compatible-EMRS}. 
Si spacers and cap were grown at the temperature of {530\textcelsius} in heterostructures with thick spacers ($h_{\rm Si}$ = 50~nm) \cite{classification, CMOS-compatible-EMRS,VCIAN-2012} and at {360\textcelsius} in the structures with thin spacers ($h_{\rm Si}$ = 5 and 8~nm)\cite{Yur_JNO}.

HR~TEM images of the grown structures were obtained using a Libra-200 FE HR instrument (Carl Zeiss) after the ion treatment in argon plasma using a Model 1010 ion mill (E.~A.~Fischione Instruments) for obtaining thin defectless lamellae. Images with the highest contrast of separate atoms were selected among all the obtained ones; they were processed using the authors' software that applied the peak pairs algorithm for analysing lengths of separate inter-atomic bonds (Fig.\,\ref{fig:Colors}) \cite{Peak_Pairs}. An asymmetrical averaging by 2 neighbouring bonds in the investigated direction and by 5 neighbouring bonds in the perpendicular direction was used in order to eliminate single errors in bond lengths. The resulting confidence band of coloring is shown by yellow dashed lines in final images at which it can be applied to the area of strained bonds. 

Raman spectra were obtained using a LabRAM HR Evolution confocal scanning Raman microscope (Horiba Scientific). 
Measurements were carried out using linearly polarized excitation at a wavelength of 632.8 nm, 1800 lines/mm diffraction grating (spectral resolution is 1 cm$^{-1}$), and $\times$100 objective (numerical aperture $A_{\rm N} = 0.90$).
The spot size was $\sim 0.43~\mu$m that corresponds to $\sim 10^{2}$ excited QDs for samples having Ge layers of 6~\AA~and $\sim 10^{3}$ excited QDs for samples having Ge layers of 10~\AA.
We used unpolarized light detection in order to have a significant signal-to-noise ratio. 
The Raman spectra were recorded at 1.8, 0.7 and 0.35 mW incident laser emission power; the integration time was 10\,s at each point. The statistics were collected from at least 10 points from different parts of the same specimen. 
Gaussian or Voigt profiles were used for the deconvolution of Raman bands.

In addition, uncapped Ge films with quantum dots on the Si(001) surface were examined using an STM GPI-300 UHV scanning tunnelling microscope (Sigma Scan) linked with the MBE chamber via a UHV transfer line \cite{classification}.

STM and some HR~TEM images were processed using the WSxM software pack \cite{WSxM}.


\section{\label{sec:results}Results }

\subsection{\label{sec:TEM}High resolution TEM and peak-pairs analysis}

\begin{figure*}[ht]
	\centering
	\includegraphics[width=.32\textwidth]{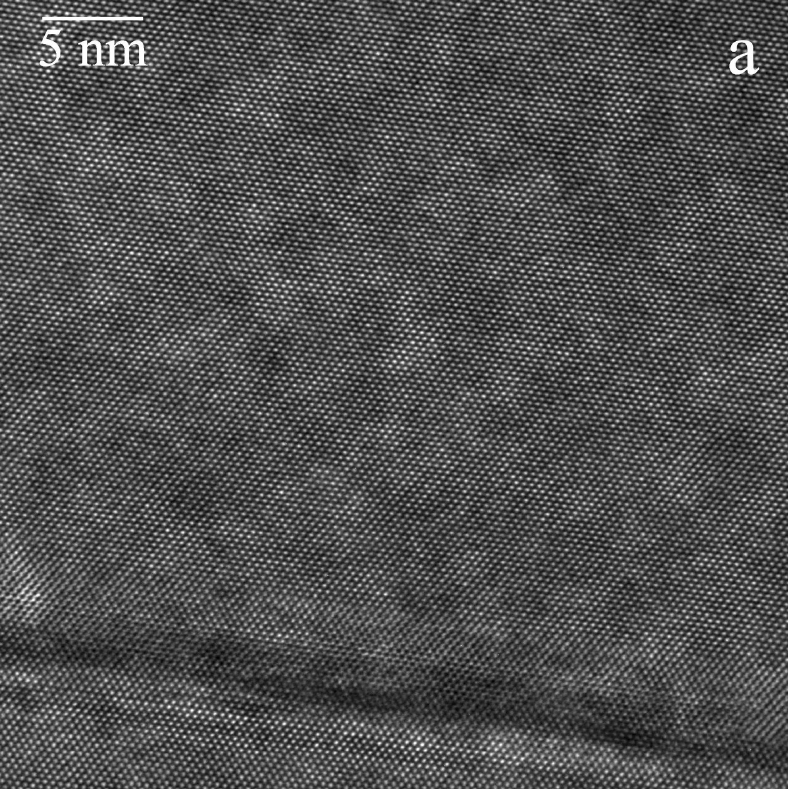} 
	\includegraphics[width=.32\textwidth]{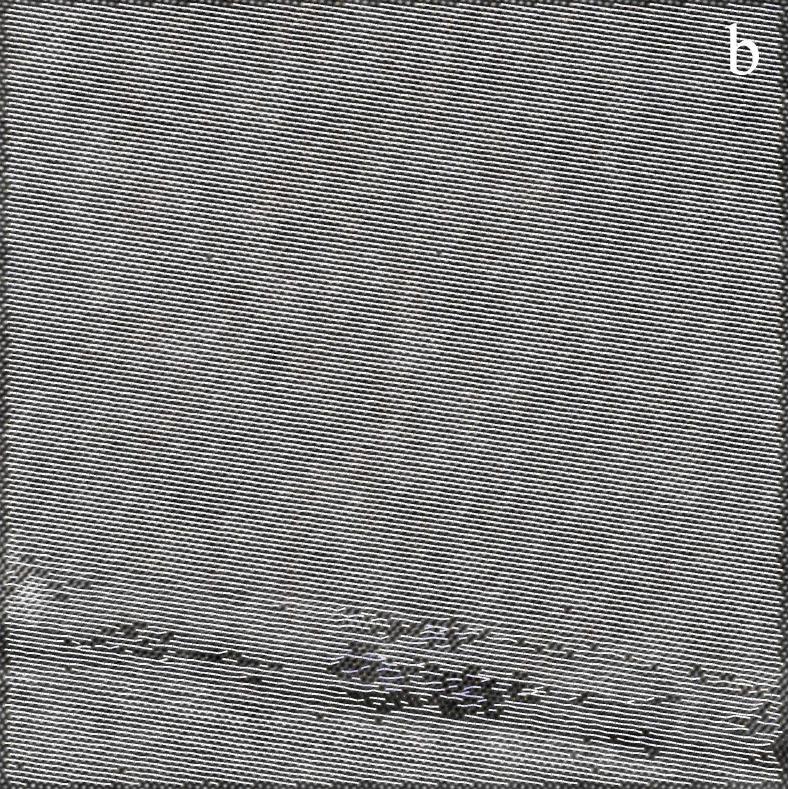} 
	\includegraphics[width=.32\textwidth]{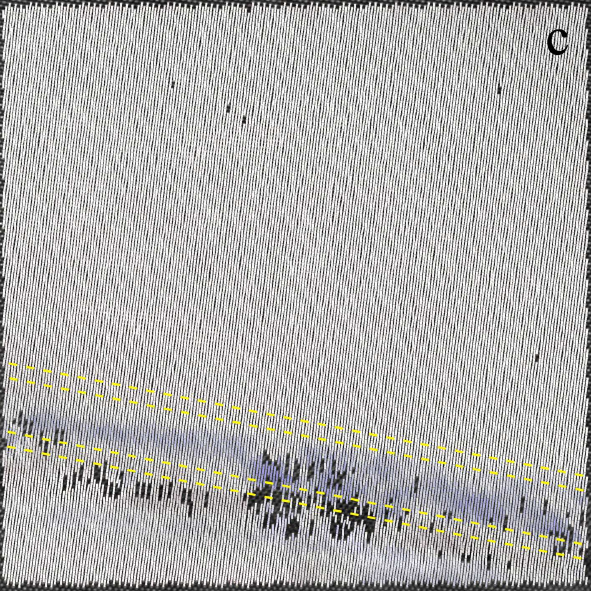}
	\caption{\label{fig:10A-1}
		A HR~TEM image 
		(a) a single quantum dot in a structure consisting of five layers of Ge/Si(001) QDs in Si matrix
		(the Ge coverage 
		{$h_{\rm Ge} =10$~\AA}, 
		the Ge deposition temperature 
		{$T_{\rm Ge} = 360$\textcelsius},
		the Si spacers thickness
		{$h_{\rm Si} \approx500$~\AA},
		the Si deposition temperature 
		{$T_{\rm Si} = 530$\textcelsius}) 
		and the results of image processing in [110]~(b) and [001]~(c) directions using the peak pairs algorithm; 
		the palette of correspondence between strain values of interatomic bonds and colour is given in Fig.~\ref{fig:Colors}. The domain with increased lattice parameter is limited by yellow dashed lines; the distance between neighbouring lines corresponds to the confidence band.
	}
\end{figure*}  

\begin{figure*}[ht]
	\centering
	\includegraphics[width=.9\textwidth]{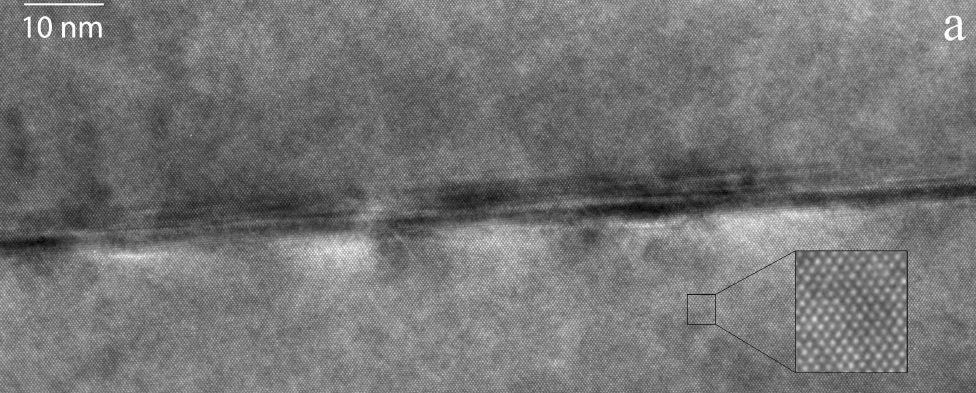}\\
	\includegraphics[width=.9\textwidth]{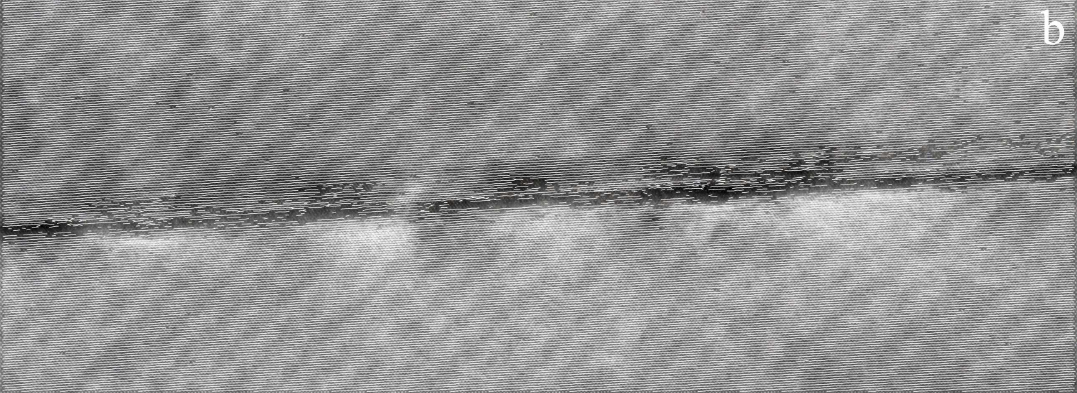}\\
	\includegraphics[width=.9\textwidth]{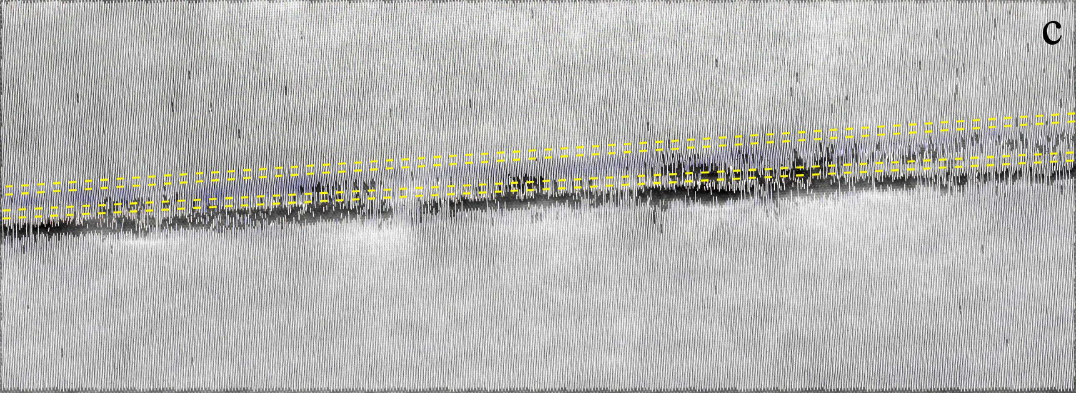}
	\caption{\label{fig:10A-2}
		A HR~TEM image
		(a) of a fragment of a multilayer structure with Ge/Si(001) QD arrays in Si matrix 
		(the fragment demonstrates only one layer of the five-layer structure; 
		{$h_{\rm Ge} =10$~\AA}, 
		{$T_{\rm Ge} = 360$\textcelsius}, 
		{$h_{\rm Si} \approx500$~\AA},
		{$T_{\rm Si} = 530$\textcelsius})
		and the results of image processing in [110] (b) and [001] (c) directions using the peak pairs algorithm;
		for the correspondence between strain values and colour, see Fig.~\ref{fig:Colors}. The domain with increased lattice parameter is limited by yellow dashed lines; the distance between neighbouring lines corresponds to the confidence band.
	}
\end{figure*}

Fig.\,\ref{fig:10A-1} demonstrates a HR~TEM image of a single quantum dot in the sample consisting of 5 layers of Ge QDs grown by depositing of {10~\AA} of Ge at the temperature of {360\textcelsius} separated by Si layers of 50 nm thick grown at the temperature of 530{\textcelsius} and the results of its processing using the peak-pairs algorithm in [110] and [001] directions.
A narrow crystal region, $\sim10$ to 15 monolayers in thickness,\footnote{%
The thickness of 1 monolayer of Si is $\sim1.36$~\AA.}
 is observed above the layer of QDs; the lattice parameter in this region is increased by 2 to 4~\% relative to the unstrained Si crystal in the growth direction ([001]).
In the [110] direction, in the crystal growth plane, the lattice over QD layer, like in the whole presented domain, corresponds to the unstrained Si crystal. The obtained result cannot be assigned to peculiarities of sample preparation for TEM because lamella distortions caused by high strains must spread through the whole structure (it can be observed at Fig.\,\ref{fig:10A-Super}a,\,b) while the possible Ge extruding in the direction of sample thinning cannot influence on strains in the visualized plane. Negligible strain fluctuations are observed throughout the whole studied region. They are likely artefacts occurring due to the lamella minor local bending. Uncoloured dark grey spots are present in the maps due to absence of atomic contrast in the corresponding  parts of the HR~TEM image. 


An image of this structure fragment including several QDs is demonstrated in Fig.\,\ref{fig:10A-2}a. The results of its processing in the [110] and [001] directions are similar to those obtained for the image of a single QD (Fig.\,\ref{fig:10A-2}b,\,c): 
the lattice is stretched by 2 to 4~\% compared to unstrained Si in the [100] direction in the 10 to 15 monolayers thick domain above the Ge QD layer and remains unchanged in the [110] direction. 

Fig.\,\ref{fig:6A-1} depicts an image of a structure obtained by depositing five {6~\AA} thick layers of Ge at the temperature of {360\textcelsius} separated by 50~nm thick Si layers grown at the temperature of 530{\textcelsius} and the results of its processing in [110] and [001] directions using the pick-pairs algorithm.
No strains of the crystal lattice are observed in this sample either in the [110] direction or in the [001] one.

\begin{figure}[ht]
	\centering
	\includegraphics[width=.32\textwidth]{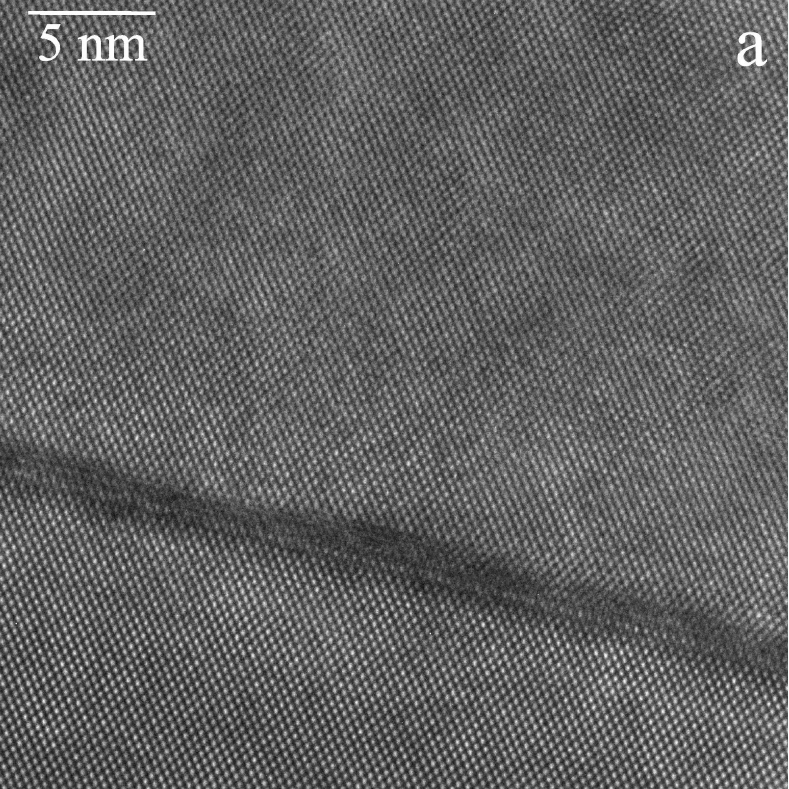} 
	\includegraphics[width=.32\textwidth]{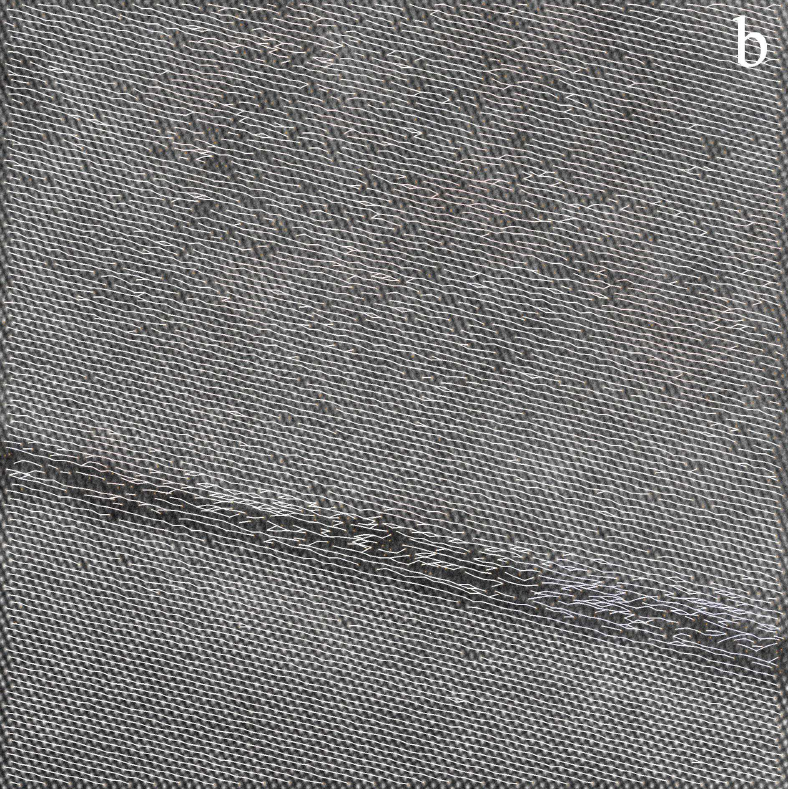} 
	\includegraphics[width=.32\textwidth]{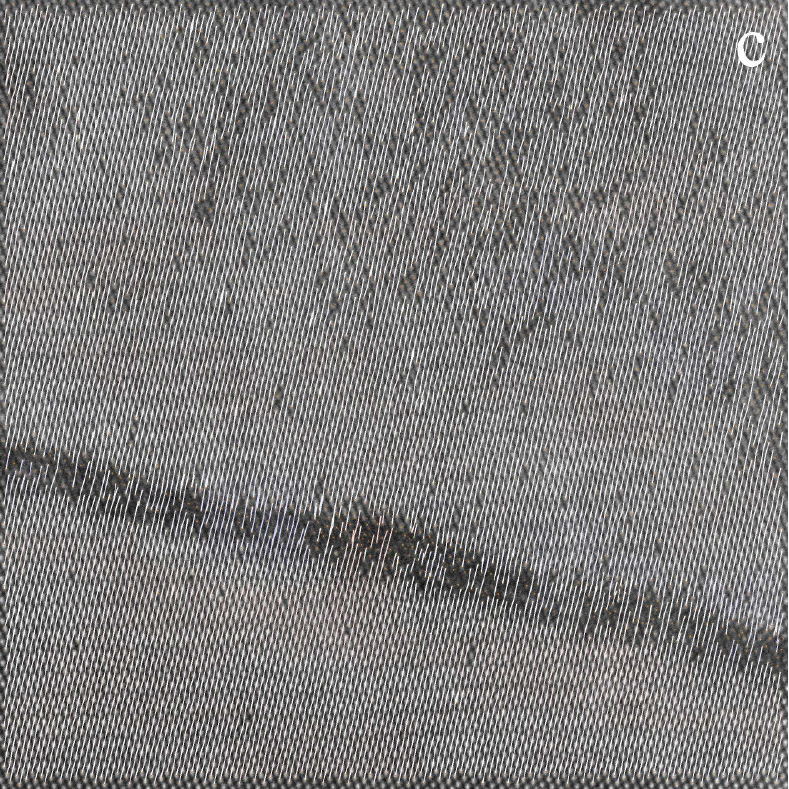}
	\caption{\label{fig:6A-1}
		A HR~TEM image 
		(a) of a fragment of a multilayer structure with Ge/Si(001) QD arrays in Si matrix
		({$h_{\rm Ge} =6$~\AA}, 
		{$T_{\rm Ge} = 360$\textcelsius},
		{$h_{\rm Si} \approx500$~\AA},
		{$T_{\rm Si} = 530$\textcelsius}) 
		and the results of image processing in [110] (b) and [001] (c) directions using the peak pairs algorithm;
		crystal lattice strains are observed neither in [110] nor in [001] direction.
	}
\end{figure}

When a thin layer of single-crystalline Si is deposited on a Ge crystal, the Si crystal lattice is known to stretch in the growth plane due to greater lattice parameter of Ge \cite{Si/SiGe-MOSFET_Review}.
This effect is widely used in manufacturing of semiconductor electronic devices for obtaining of Si layers with the enhanced carrier mobility \cite{Strained_Si_SiGe_Ge_FET}.
However, no stretching of Si lattice in the growth plane near the Ge layer is observed in the samples with thin Ge layers located between thick Si layers studied by us (Figs.\,\ref{fig:10A-1}b, \ref{fig:10A-2}b and \ref{fig:6A-1}b).
We assume that the observed part of a crystal with the increased lattice parameter in the growth direction [001] corresponds to the domain of a mixed composition Si$_x$Ge$_{1-x}$, the lattice parameter of which linearly depends on $x$ \cite{Phonon_Spectra_GeSi}.
According to the Vegard's law \cite{Vegard_law}, an increase in the lattice parameter by 2 to 4~\% corresponds to the Ge content from $\sim 50$ to 100~\%.
Thus, it might be supposed that the stresses over a layer of Ge QDs occurring due to the further growth of a thick Si layer are relieved by compressing the thin domain of a mixed Si$_x$Ge$_{1-x}$ composition in the [110] direction with preservation of the own lattice parameter of Si$_x$Ge$_{1-x}$ in the [001] direction.


\subsection{\label{sec:Raman}Raman spectroscopy}

\begin{figure}[htp]
	\centering
	\includegraphics[width=.98\textwidth]{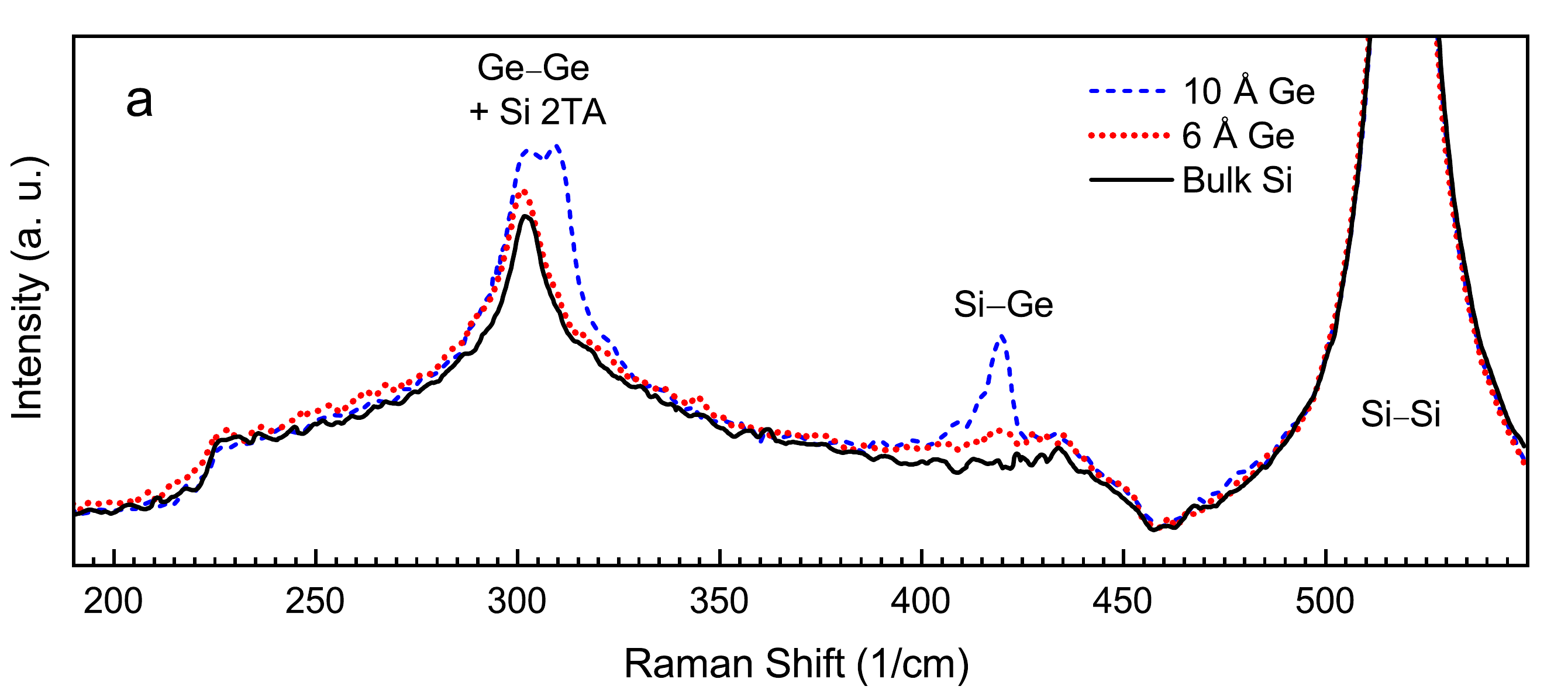}\\
	\includegraphics[scale=.6,width=.49\textwidth]{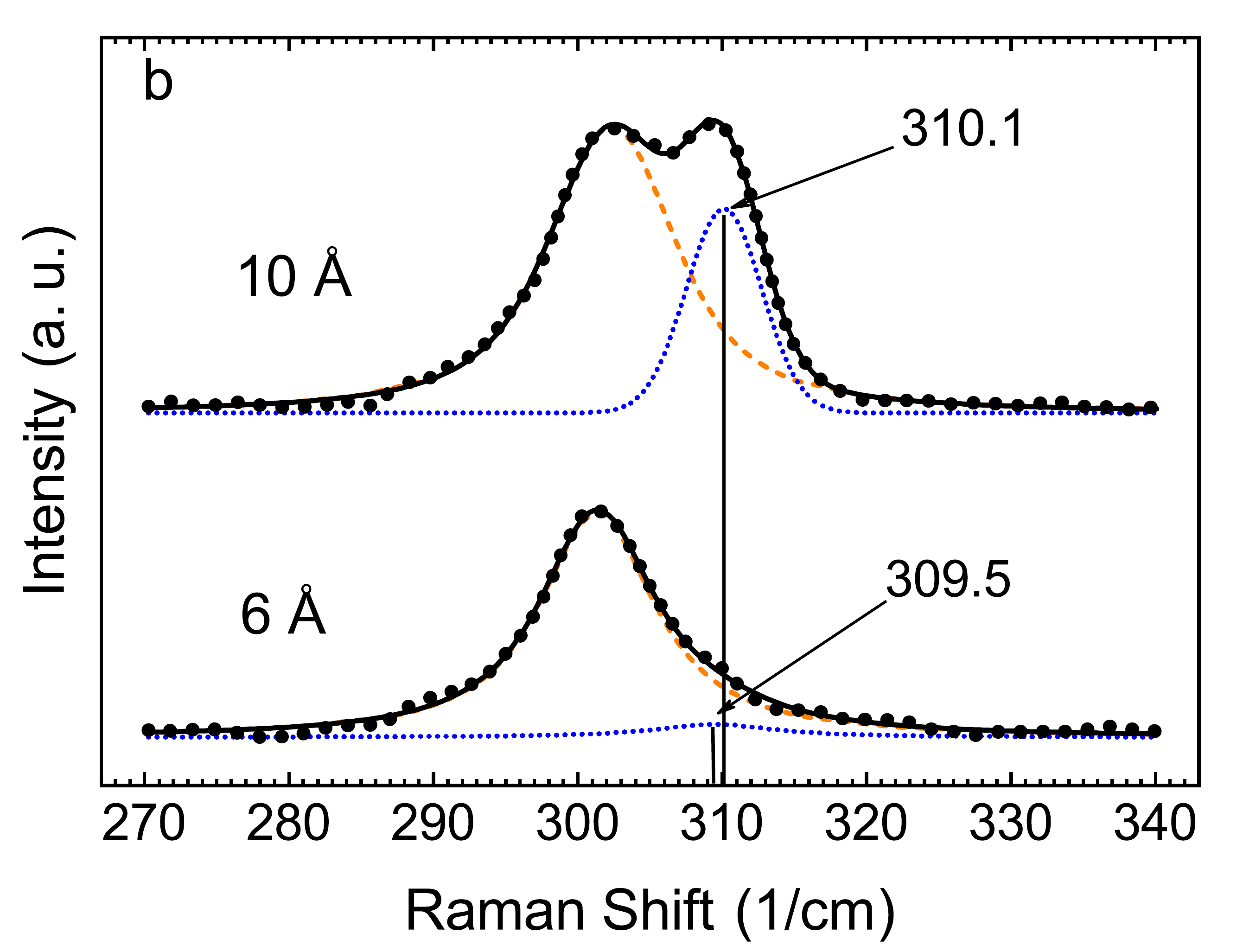}
	\includegraphics[scale=.6,width=.49\textwidth]{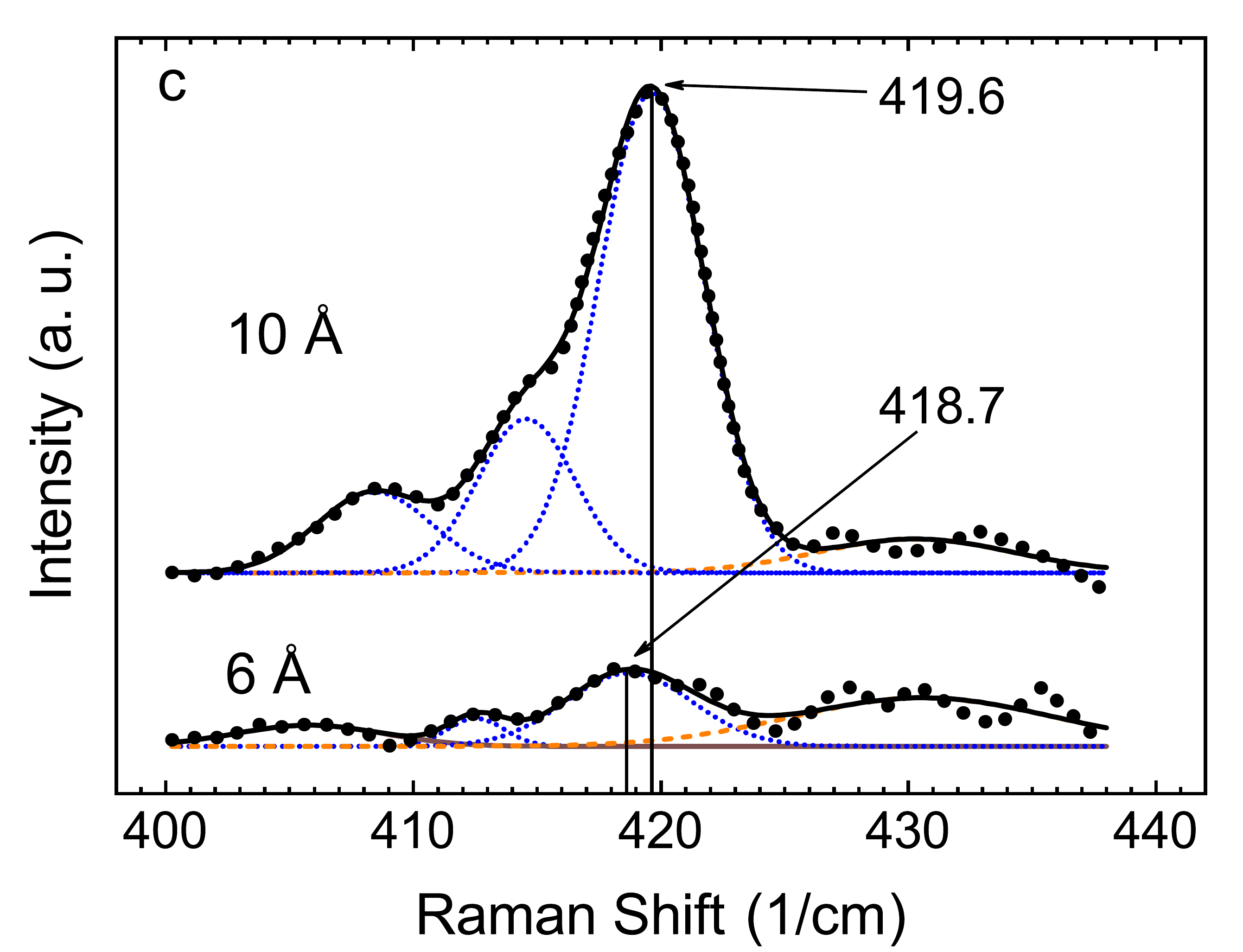}
	\caption{\label{fig:Raman-6&10A_Ge}%
		Survey Raman spectra (a) of Ge/Si heterostructures with five Ge layers of different thickness (%
		{$h_{\rm Ge} =6$ and 10~\AA},
		{$T_{\rm Ge} = 360$\textcelsius},
		{$h_{\rm Si} \approx500$~\AA}, 
		{$T_{\rm Si} = 530$\textcelsius}%
		) 
		and 
		the detailed spectra in the vicinity of the 
		Ge--Ge (b) 
		and 
		Si--Ge (c) 
		bands 
		(%
		the experimental data are depicted by the black filled circles, 
		the black solid lines depict the cumulative results of the band deconvolution, 
		the peaks associated with the Si--Si bond vibrations are depicted by the orange dashed lines 
		and 
		the peaks associated with the 
		Ge--Ge 
		and 
		Si--Ge 
		bonds vibrations are depicted by the blue dotted lines);
		arrows and numerical values indicate the spectral positions of the studied lines obtained by the deconvolution of a corresponding band.
	}
\end{figure}

\begin{figure}[ht]
	\centering
	\includegraphics[width=.98\textwidth]{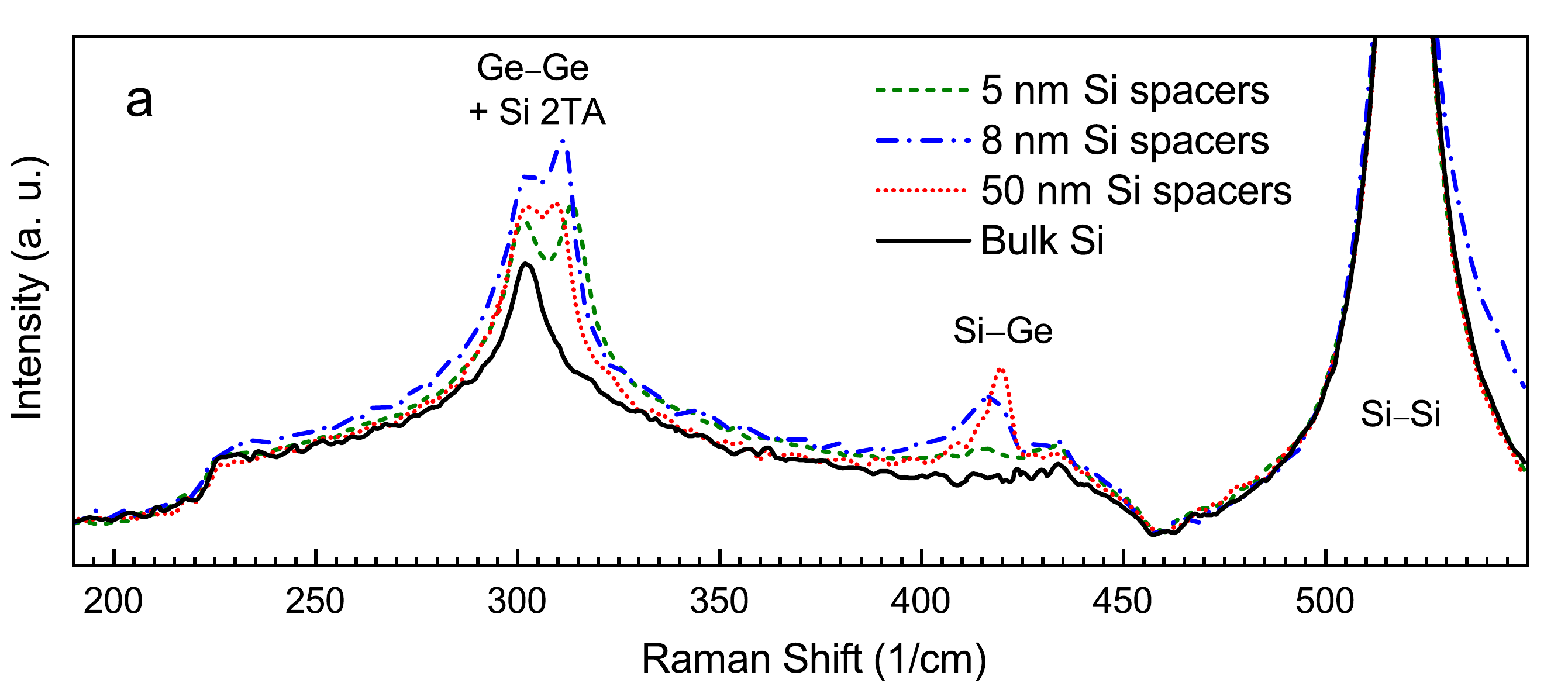}\\
	\includegraphics[width=.49\textwidth]{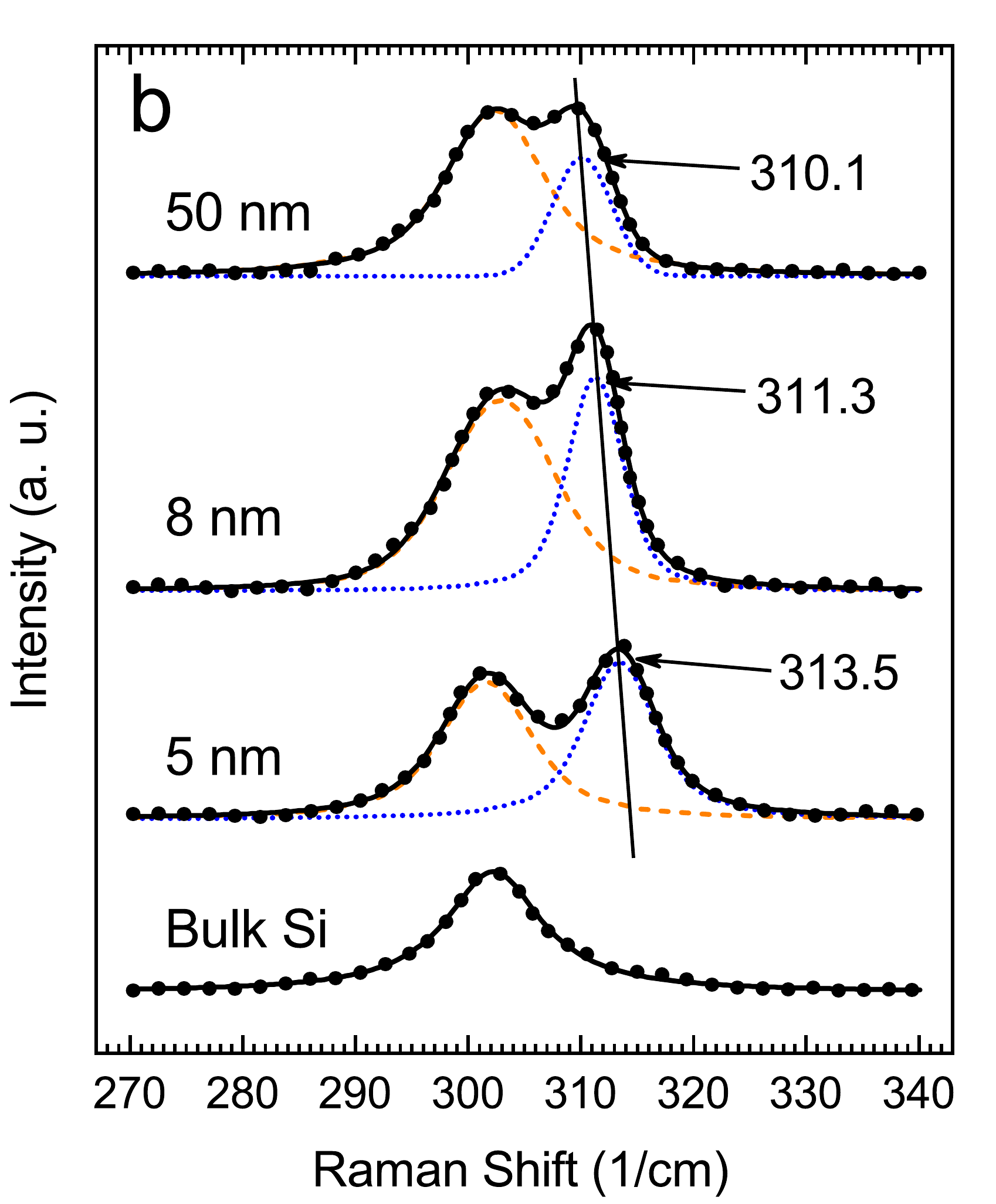}
	\includegraphics[width=.49\textwidth]{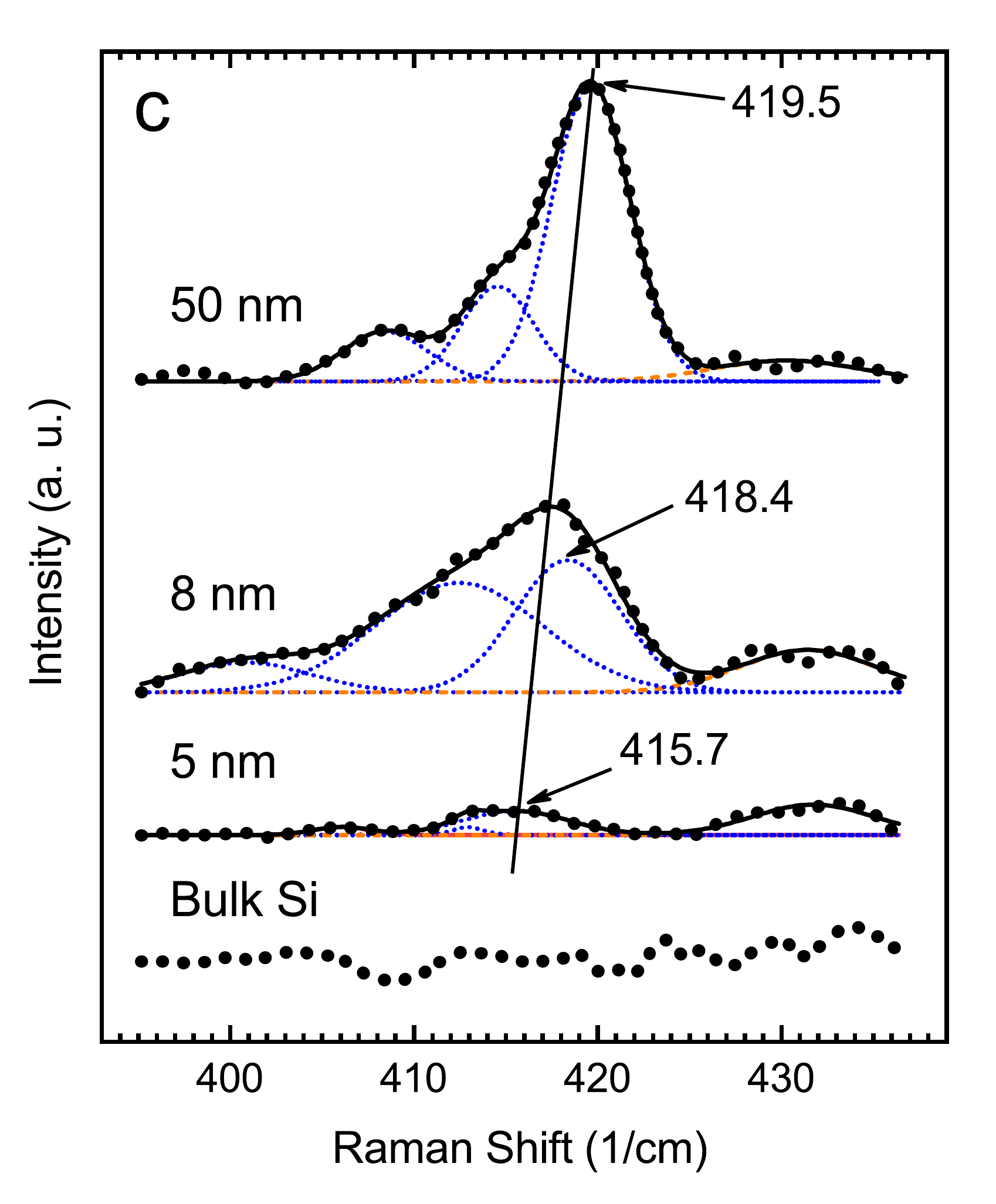}
	\caption{\label{fig:Raman-spacers}%
		Survey Raman spectra (a) of Ge/Si heterostructures with five Ge layers separated by Si spacers of different thickness (%
		{$h_{\rm Ge} =10$~\AA},
		{$T_{\rm Ge} = 360$\textcelsius},
		{$h_{\rm Si} = 50$, 80 and $\sim500$~\AA}, 
		{$T_{\rm Si} = 360$\textcelsius}
		for {$h_{\rm Si} = 50$ and 80~\AA}
		and
		{530\textcelsius}
		for {$h_{\rm Si} \approx 500$~\AA}%
		) 
		and 
		the detailed spectra in the vicinity the
		Ge--Ge (b) 
		and 
		Si--Ge (c) 
		bands 
		(%
		the experimental data are depicted by the black filled circles, 
		the black solid lines depict the cumulative results of the band deconvolution, 
		the peaks associated with the Si--Si bond vibrations are depicted by the orange dashed lines 
		and 
		the peaks associated with the 
		Ge--Ge 
		and 
		Si--Ge 
		bonds vibrations are depicted by the blue dotted lines);
		arrows and numerical values indicate the spectral positions of the studied lines obtained by the deconvolution of a corresponding band.
	}
\end{figure}

To verify the above suggestions about the presence and the role of the mixed Si$_x$Ge$_{1-x}$ layer, Raman scattering spectra of structures with Ge QDs of different size separated by Si layers of different thickness were analysed. 
Fig.~\ref{fig:Raman-6&10A_Ge} demonstrates Raman spectra of the samples investigated by HRTEM with the Ge coverage 6 and 10 \r{A} that represent spectra typical for Ge/Si structures. 
The strong band assigned to the Si--Si bond vibration peaked at $\sim\,$520~cm$^{-1}$, the peak assigned to the Ge--Ge bond vibration in the range from 300 to 320~cm$^{-1}$, the peak of the 2TA phonon of Si at $\sim\,$302~cm$^{-1}$ and the line arising due to the Si--Ge bond vibration peaked in the range from 410 to 430~cm$^{-1}$ are registered in the spectra. 
The detailed spectra of the regions near the bands of the Ge--Ge and Si--Ge vibrations are shown in panels b and c of Fig.~\ref{fig:Raman-6&10A_Ge}, respectively. 
We do not subtract the contribution of the Si--Si bond vibration line wings from the spectra during the analysis of the Ge--Ge bond vibration since the peak related to the Si--Si line in the reference sample and that in the etched part of the heterostructure may somewhat differ in their width and fine structure from the peak of the Si--Si bond vibration in the sample with embedded QDs. 
However, the deconvolution by elementary peaks was applied to reveal the true position of the Ge--Ge vibration band especially veiled by the signal from 2TA phonon of Si in the sample with low amount of deposited Ge. 
The red shift of the Ge--Ge line for the sample with 6 \r{A} of deposited Ge relative to sample with 10 \r{A} of deposited Ge is clearly observable that corresponds to slightly less strained Ge (Fig.~\ref{fig:Raman-6&10A_Ge}b). 
At the same time, the Si--Ge related band almost vanishes in the sample with 6 \r{A} of deposited Ge (Fig.~\ref{fig:Raman-6&10A_Ge}c), which means that the total number of Si--Ge bonds in the crystal is very low that correlates with the HRTEM analysis.

Si--Ge bands (Fig.~\ref{fig:Raman-6&10A_Ge}c and \ref{fig:Raman-spacers}c) and Ge--Ge bands of some samples (Fig.\ref{fig:Raman-spacers}b) are composed of several spectral peaks. 
This indicates that Ge and the domain of a mixed composition consist of different phases of varying strain. 
There is no contradiction with our suggestion in this observation.
However, it requires a special detailed investigation.


\subsection{\label{sec:Chains}Long chains of Ge quantum dots}

We have also investigated multilayer Ge/Si(001) heterostructures containing long chains of Ge QDs.

\begin{figure}[ht]
	\centering
	\includegraphics[width=.4\textwidth]{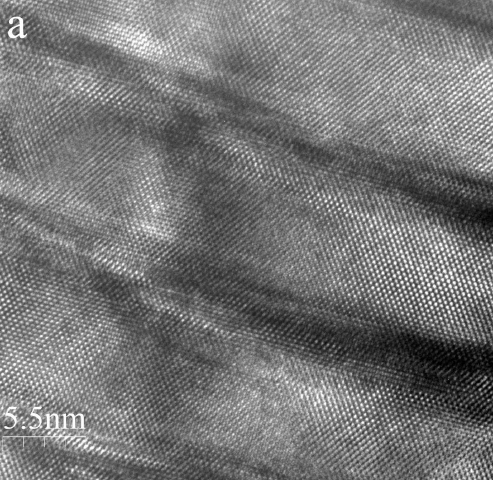} 
	\includegraphics[width=.4\textwidth]{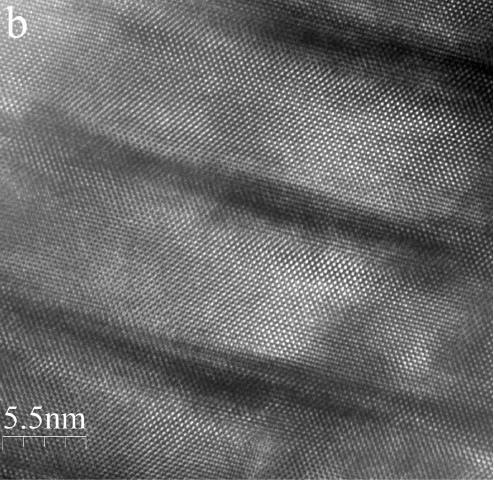} \\
	\includegraphics[width=.4\textwidth]{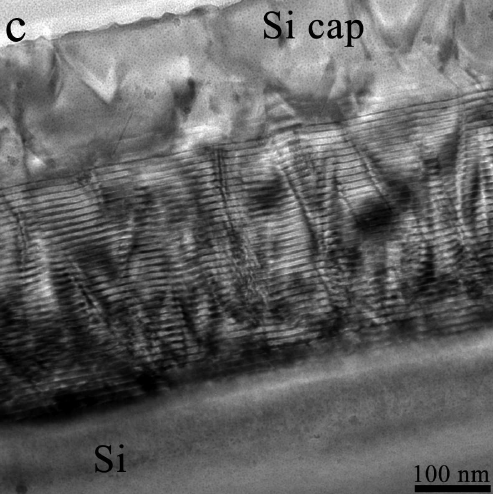}
	\includegraphics[width=.4\textwidth]{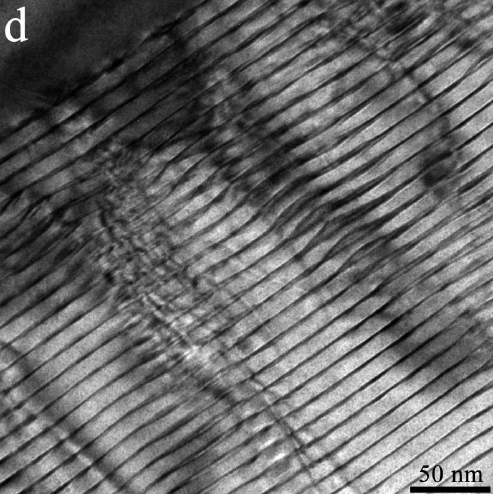}
	\caption{\label{fig:10A-Super}%
		Typical HR~TEM images of Ge/Si(001) heterostructures composed by 30 layers of Ge QD arrays separated by thin Si spacers
		({$h_{\rm Ge} =10$~\AA}, 
		{$T_{\rm Ge} = 360$\textcelsius},
		$h_{\rm Si} \approx100$~{\AA}, 
		{$T_{\rm Si} = 360$\textcelsius}).
	}
\end{figure}

Fig.\,\ref{fig:10A-Super} demonstrates  HR~TEM images of Ge/Si(001) heterostructures composed by 30 layers of Ge QD arrays separated by thin Si spacers.
The distribution of the stress field in the heterostructure is seen in them in numerous details. 

Unfortunately, we failed to analyse the HR~TEM images of multilayer structures with thin spacers ($h_{\rm Si}< 10$~nm) between QD layers and long chains of Ge QDs using the peak pairs algorithm. 
Images of such structures obtained with atomic resolution were taken at high magnification and displayed areas of small sizes near the Ge QD layers (Fig.\,\ref{fig:10A-Super}a,\,b).
The atomic contrast in them changed from positive to negative in different areas because of the slight bending of the lamella due to the internal stress embedded in these heterostructures.
It was impossible to select a lattice image in the micro images of regions of small sizes that could be taken for sure as an image of an undistorted Si lattice, i.e. it was impossible to analyse such images using the peak pairs algorithm.

\begin{figure}[ht]
	\centering
	\includegraphics[width=.5\textwidth]{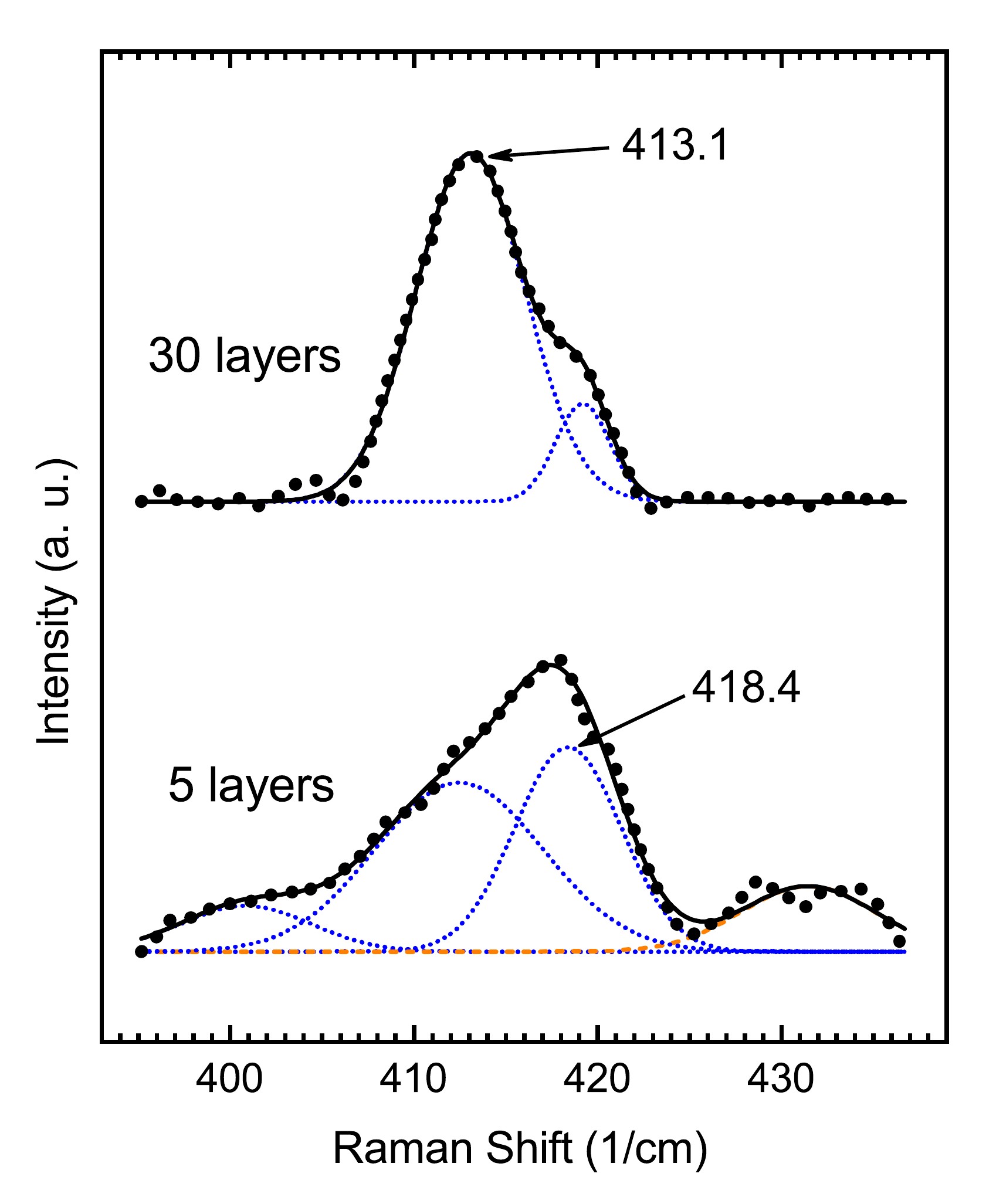} 
	\caption{\label{fig:Raman_10A-Super}%
		Raman spectra of Ge/Si heterostructures containing 30 and 5 Ge layers
		in the vicinity of the bands assigned to the Si--Ge vibration 
		({$h_{\rm Ge} = 10$~\AA}, 
		{$T_{\rm Ge} = 360$\textcelsius},   
		$h_{\rm Si} \approx 80$~{\AA} for the 5-layer structure 
		and 
		100~{\AA} for the 10-layer structure, 
		{$T_{\rm Si} = 360$\textcelsius}).
		The experimental data are depicted by the filled black circles, the solid black lines depict the cumulative result of the deconvolution, the peaks associated with the Si--Si vibrations are shown by orange dashed lines and the peaks associated with the Si--Ge vibrations are presented by blue dotted lines; arrows and numerals indicate the spectral positions of the studied lines obtained by the deconvolution of the corresponding bands.
	}
\end{figure}

However, the obtained Raman spectra of the Ge/Si(001) heterostructures with thin Si spacers (Fig.~\ref{fig:Raman_10A-Super}) give a sufficient information enabling the explanation of the correlated growth of Ge QD chains. 
The spectra obtained at the structure composed of 30 layers of Ge QDs separated by Si spacers of 10~\r{A} in thickness demonstrate a much greater red shift of the band assigned to the Si--Ge bond vibrations than those obtained at the structure composed of 5 layers of Ge~QDs, i.e. the more layers in a Ge/Si(001) heterostructure with QD chains the greater is the red shift. 
Since the Si--Ge related band is composed of several spectral peaks, the mixed composition domain consists of differently strained phases again, yet the stretched phase predominates in this case.  


\section{\label{sec:discuss}Discussion}

There are two fundamental mechanisms of intermixing between Si and Ge on the heterointerface. 
The first one is the well-known effect of Ge segregation during deposition of Si on Ge that takes place even at low temperatures starting from $\sim200${\textcelsius} \cite{47_Atomistic_interfacial_mixing_Si-Ge,48_SiGe_interfacial_ordering_Jesson,49_Ge_segregation@Si-Ge_stepped,50_Ge_segregation@Si-SiGe_interfaces,51_Role_Ge_segregation_Si-Ge_ordering,52_Step-driven_lateral_segregation_SiGe,53_Segregation@Si-Ge-Si}. 
The second one is the diffusion of Ge atoms through the Si lattice, which is significantly suppressed at the temperature we used in our experiment but can also be stimulated by stresses in the crystalline lattice. 
Segregation and temperature-induced diffusion of Ge must be present in all samples despite of the thickness of Ge and Si layers. 
Fig.~\ref{fig:Raman-spacers} demonstrates Raman spectra of samples containing 5 layers of Ge QDs grown by depositing of 10~\r{A} of Ge at the temperature of 360{\textcelsius} and separated by Si layers of various thicknesses. 
One can see from the plot (Fig.~\ref{fig:Raman-spacers}c) that the intensity of the Si--Ge related band extremely grows up for the sample with Si spacers of 8~nm in comparison with the sample with Si spacers of 5~nm followed by the red shift of the Ge--Ge peak by $\sim2$~cm$^{-1}$ (Fig.~\ref{fig:Raman-spacers}b) that corresponds to decreasing of strains in Ge. At the same time the blue shift of the Si--Ge peak definitely stands for the accumulation of strains in the domain of a mixed composition. 
This result can be described in the terms of stress-induced diffusion, which becomes significant at the sufficient thickness of the Ge layer accompanied with thick enough Si coverage.

We suppose, that the diffusion of Ge into the Si layer resulting in the formation of the mixed composition domain the most actively occurs from the \{105\} facets of Ge huts after the growing Si layer reaches some critical thickness, at which the stress at the Si/Ge interface becomes sufficient to launch the accelerated Ge diffusion into Si.
The diffusion should stop as soon as a mixed layer of Si$_x$Ge$_{1-x}$ is formed, providing the stress relaxation.
An important point is that the most significant role is played by stresses arising due to the difference in the lattice parameters of Si and Ge near the \{105\} faces of huts, and especially at the edges of monoatomic steps \cite{atomic_structure}, since the work function of the Ge atoms at the step edge should be significantly lower than on the (001) plane \cite{Free_energies_Ge(001),49_Ge_segregation@Si-Ge_stepped,52_Step-driven_lateral_segregation_SiGe}. 
Additionally, a breaking barrier for loaded interatomic Ge--Ge bonds drastically decreases compared to that for unloaded ones \cite{Atomic-Level_Fracture_Solids}, and the higher the applied stress the lower the breaking barrier.
Thus, the probability of detachment of Ge atoms and their hopping into the Si lattice is the highest namely at steps of the \{105\} facets of huts where the binding energy of Ge--Ge bonds is lowered and the bonds are highly loaded because of the difference in the lattice parameters of Si and Ge.
(Note, that the Si--Si bonds are also highly loaded in the nearest vicinity of steps of the \{105\} facets, and their breaking barrier is significantly lowered that should increase the likelihood of Ge atoms hopping in the Si lattice as well as the probability of the Ge--Si bonds formation.)

\begin{figure}[ht]
	\centering
	\includegraphics[width=.4\textwidth]{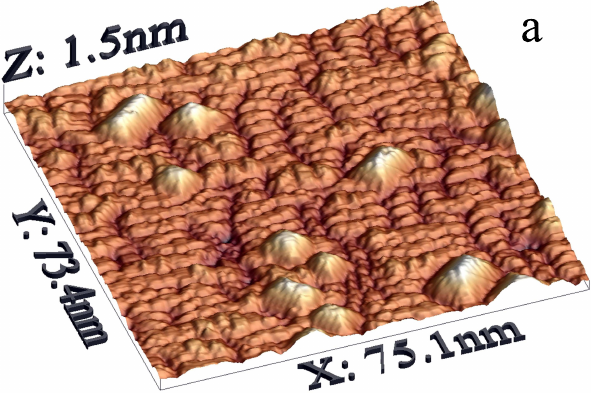}~~
	\includegraphics[width=.4\textwidth]{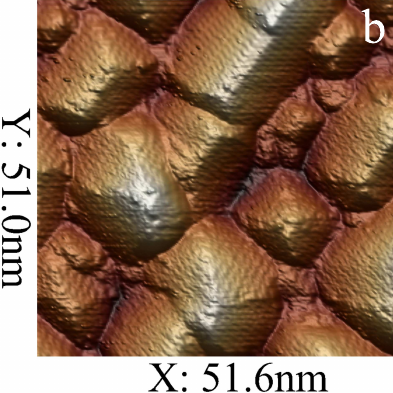}
	\caption{\label{fig:STM}%
		STM images of Ge hut cluster arrays on Si(001) for
		(a)~{$h_{\rm Ge} =6$~\AA}
		and
		(b)~{$h_{\rm Ge} =10$~\AA}
		({$T_{\rm Ge} = 360$\textcelsius}).
	}
\end{figure}

Huts that are present in structures with low thickness of the Ge layers ($h_{\rm Ge} = 6$ to 7 \AA) occupy only a small part of the surface area of the wetting layer (Fig.~\ref{fig:STM}a), and they themselves are small and their \{105\} facets do not contain enough Ge atoms to provide any significant intermixing of Si and Ge at the interface \cite{classification,VCIAN-2012,CMOS-compatible-EMRS,Yur_JNO,Growing_Ge_hut-structure,initial_phase,VCIAN2011}. 
At the same time Raman spectra analysis shows (Fig.~\ref{fig:Raman-6&10A_Ge}b) that segregation of Ge providing intermixing within 1 to 2 monolayers \cite{47_Atomistic_interfacial_mixing_Si-Ge,48_SiGe_interfacial_ordering_Jesson,49_Ge_segregation@Si-Ge_stepped,50_Ge_segregation@Si-SiGe_interfaces},  
which is not resolved in our HRTEM images, is enough to keep wetting layers partially relaxed. At thicker layers of Ge ({$h_{\rm Ge} > 7$~\AA}) a fraction of the wetting layer surface that is not occupied by huts decreases and at Ge coverages {$h_{\rm Ge} > 9$~\AA} already almost all wetting layer is covered with huts (Fig.~\ref{fig:STM}b) \cite{classification,VCIAN-2012,CMOS-compatible-EMRS,Yur_JNO,VCIAN2011}. 
Thus Si is deposited only on \{105\} facets of huts at Ge coverage {$h_{\rm Ge} =10$~\AA} and a process of Ge and Si stress-induced intermixing occurs efficiently when the thickness of overlying Si is enough to produce considerable stresses. 
It is vital to emphasize that the mixed layer is compressed in the [110] direction by the thick upper layer of Si yet keeps its lattice parameter in the [001] direction controlled by the ratio of Ge and Si fractions in the solid solution. 
Further growth of the scattering intensity from the Si--Ge bond vibration and the red shift of the Ge--Ge band for the structure with 50~nm thick Si layers grown at the higher temperature of 530{\textcelsius} is insignificant (Fig.~\ref{fig:Raman-spacers}) and is associated rather with temperature-induced diffusion.%

Similar results were previously obtained by us by analysing Fourier transforms of HRTEM images of small regions of Si spacers and caps directly adjacent to the apexes of Ge huts \cite{VCIAN-2012,Yur_JNO}: 
in those regions lattice periods were $\sim5.4$ and 3.8~\r{A} along the [001] and [110] directions in the structures with Ge coverage of 6~\r{A} that corresponds to unstrained Si, whereas in the structures with Ge coverages of 9 and 10~\r{A} they reached $\sim5.6$~\r{A} along the [001] direction that is close to the lattice period of unstrained Ge. 
Si and Ge intermixing between a Ge QD layer and the underlying Si layer did not happen in any of the samples examined in those works that is also entirely explained within the model proposed in the current article.

Note also that the discussed result completely agrees with our previous studies of Ge/Si(001) structures with Ge QDs using Raman spectroscopy \cite{our_Raman_en}: the peaks assigned to the Si--Ge vibration were absent in the Raman spectra of uncapped single-layer samples and practically absent in the spectra of any samples with {$h_{\rm Ge} =6$~\AA}.
Also, they are fully in agreement with the studies of effect of a cap on the composition of large Ge QDs 
\cite{Strain&composition_Ge/Si}.
It is noteworthy that the crystal lattice deformation was not observed in Si layers between high-temperature QDs completely composed by the Si$_x$Ge$_{1-x}$ alloy either \cite{Lattice_deformation@Ge-QD}.
This proves once again that it is more profitable for the stress to relax via the formation of the intermixing region, rather than spread into Si, which is expected due to the lower energy of the inter-atomic bonds in Si$_x$Ge$_{1-x}$ compared to pure Si \cite{J_Chem_Soc_Pak}.

The above discussion also helps explain the vertically corrected growth of Ge/Si(001) heterostructures with dense chains of Ge QDs \cite{Yur_JNO}. Long chains of Ge QDs are seen to form in the growth direction (Fig.\,\ref{fig:10A-Super}) in the multilayer structures with thin spacers ($h_{\rm Si}\lesssim 10$~nm) between QD layers. 
Local lamella bends distorting HR~TEM images and changing the contrast of images of atomic columns (Fig.\,\ref{fig:10A-Super}a,\,b), indicate the presence of high stress in the silicon layers between QDs and its accumulation during the growth of the structure with a great number of Ge layers (this is presented in numerous details in Refs.~\cite{Yur_JNO}). 
Raman spectra of the structure containing 30 layers of Ge QDs separated by 10~nm thick Si spacers demonstrate a strong red shift of the Si--Ge band in comparison with the structure containing 5 layers of Ge~QDs (Fig.~\ref{fig:Raman_10A-Super}). 
The red shift indicates that the lattice parameter of the domain of mixed composition is closer to Ge so it becomes greater during the growth of the multilayer structure, i.e. the more layers of Ge the greater is the lattice parameter. 
The Si$_x$Ge$_{1-x}$ domain consists of different phases (Si--Ge band is composed of several spectral peaks) but the stretched phase predominates in this case.  
Thus stress cannot considerably relax due to the appearance of mixed transition layers in such multilayer structures and still penetrates through them on the upper growth surface. 
At the same time we have to recognize that vertical arrangement starts already at the second layer. 
It means that even tiny stretching of Si is enough to make the point above the QD preferable for starting the growth of a new QD while a high stress accumulated during the growth of the multilayer structure probably leads to bifurcations of Ge QD chains previously observed in such heterostructures \cite{Yur_JNO}.


\section{\label{sec:conclusion}Conclusion}

Summarizing the above we would like to emphasise the main points of the article.

In the studied Ge/Si(001) heterostructures, a thin domain of the crystal ($\sim 10$ to 15 monolayers thick), located over a layer of Ge quantum dots, with the lattice parameter in the [001] growth direction exceeding that of unstressed silicon was discovered by analysis of HR~TEM images using the peak pairs algorithm.
The observation is explained by the mixed Si$_x$Ge$_{1-x}$ composition of the crystal lattice in this region, which has the parameter exceeding the lattice parameter of pure unstrained Si. 
The stress above the quantum dot relaxes by compressing the Si$_x$Ge$_{1-x}$ crystal in the [110] direction, since the lattice parameter in this direction corresponds to unstrained Si.
The transition region of the mixed composition is present only above the QD layer and is absent below it. 
It is also absent in structures with thin Ge layers and small QDs, which do not cover the  most part of the wetting layer area.

We conclude that the formation of the transition region is due to the stress-induced diffusion rather than the temperature-induced diffusion or the segregation, and the source of Ge atoms is steps of \{105\} QD facets. 
The analysis of Raman scattering spectra proves this conclusion and shows that Ge starts diffusing into Si after reaching a critical thickness of the covering Si layer that is from $\sim 5$ to $\sim 8$~nm. 

The obtained result explains the vertically correlated growth of Ge QDs in Ge/Si(001) heterostructures with thin Si spacers and the formation of long dense chains of Ge QDs in them at low temperatures including bifurcations of QD chains.

We believe that results presented in the article can find application in CMOS-compatible QD-based devices and will stimulate additional interest in detailed study of the atomic structure of stressed nanoscale objects and thin films.

\ack
The Center for Collective Use of Scientific Equipment of GPI RAS supported this research via presenting admittance to its equipment.
S.M.N. acknowledges the Russian Science Foundation (project 18-19-00684).


\section*{References}



\providecommand{\newblock}{}

\end{document}